\documentclass[twocolumn,pra,aps,amsmath,amssymb]{revtex4-1}
\usepackage[latin9]{inputenc}
\usepackage{color}
\usepackage{amstext}
\usepackage{graphicx}
\usepackage[unicode=true,pdfusetitle,
 bookmarks=false,
 breaklinks=false,pdfborder={0 0 1},backref=false,colorlinks=false]
 {hyperref}
\hypersetup{
 bookmarksnumbered=false,bookmarksopen=false}

\makeatletter
\@ifundefined{textcolor}{}
{%
 \definecolor{BLACK}{gray}{0}
 \definecolor{WHITE}{gray}{1}
 \definecolor{RED}{rgb}{1,0,0}
 \definecolor{GREEN}{rgb}{0,1,0}
 \definecolor{BLUE}{rgb}{0,0,1}
 \definecolor{CYAN}{cmyk}{1,0,0,0}
 \definecolor{MAGENTA}{cmyk}{0,1,0,0}
 \definecolor{YELLOW}{cmyk}{0,0,1,0}
}

\usepackage{txfonts}

\makeatother

\begin{document}

\title{Robust Dynamical Decoupling Sequences for Individual Nuclear Spin
Addressing}

\author{J. Casanova}
\email{jcasanovamar@gmail.com}
\affiliation{Institut f\"ur Theoretische Physik, Albert-Einstein-Allee 11, Universit\"at
Ulm, D-89069 Ulm, Germany}

\author{Z.-Y. Wang}
\email{zhenyu3cn@gmail.com}
\affiliation{Institut f\"ur Theoretische Physik, Albert-Einstein-Allee 11, Universit\"at
Ulm, D-89069 Ulm, Germany}

\author{J. F. Haase}
\email{jan.haase@uni-ulm.de}
\affiliation{Institut f\"ur Theoretische Physik, Albert-Einstein-Allee 11, Universit\"at
Ulm, D-89069 Ulm, Germany}

\author{M. B. Plenio}
\email{martin.plenio@uni-ulm.de}
\affiliation{Institut f\"ur Theoretische Physik, Albert-Einstein-Allee 11, Universit\"at
Ulm, D-89069 Ulm, Germany}

\begin{abstract}
We propose the use of non-equally spaced decoupling pulses for high-resolution
selective addressing of nuclear spins by a quantum sensor. The analytical model of the
basic operating principle is supplemented by detailed numerical studies that demonstrate
the high degree of selectivity and the robustness against static and dynamic
control field errors of this scheme. We exemplify our protocol with an
NV center-based sensor to demonstrate that it enables the identification of individual nuclear
spins that form part of a large spin ensemble.
\end{abstract}
\maketitle

\section{Introduction}
The quantum control and detection of individual constituents of nuclear spin
ensembles represents a major technological challenge whose solution would enable
the realization of robust large scale quantum registers that are required for quantum
information processing as well as the observation of the structure and dynamics of
biomolecules. The solution requires both the hardware  of a physical sensor
that is capable, in principle, of detecting the minute magnetic fields emanating from
individual nuclear spins and the software, i.e. the control schemes and measurement
protocols, that can isolate the signal of a target spin from both, the other nuclear
spins in the ensemble and from environmental noise.

Recently, the exceptional properties of the nitrogen-vacancy (NV) center in diamond have
led to rapidly growing interest into its use as a quantum device. Microwave radiation can
manipulate coherently the electron spin of the NV center while its initialization and readout
can be achieved optically \cite{Doherty13,Dobrovitsky13}. These tools enable the control
of the hyperfine coupling of the NV center to nearby $^{13}$C nuclei to carry out their
detection and polarization \cite{Kolkowitz12, Taminiau12, Zhao12,London13, Mkhitaryan15}
as well as the realization of $^{13}$C-NV quantum gates \cite{Sar12,Taminiau14}.
Furthermore, NV centers can be implanted close to the diamond surface where they can
detect the signal of nuclear spins above the surface with single spin sensitivity
\cite{Muller14} suggesting the possibility of examining the structure of biomolecules
by means of carefully designed protocols \cite{Cai13, Ajoy15, Kost14}.

These applications need to achieve simultaneously the decoupling from environmental
noise while preserving the interaction with the object of interest.
This can be achieved by dynamical decoupling (DD) protocols that refocus undesired
couplings by means of control sequences that design filters \cite{Gordon2007,Cywinski2008,Kofman2001,Biercuk2011,Sousa2009} transmissive exclusively
for very specific interactions and frequencies. As small errors in DD protocols can
accumulate and exceed the environmental perturbation they are designed to protect against,
robust schemes as exemplified by the CPMG~\cite{Carr54,Meiboom58},
the XY-family of pulses \cite{Maudsley86, Gullion90} or concatenated DD \cite{Cai12}
are highly desirable. Whereas these concepts are well understood for protecting single
isolated qubits, their extension to multiple and interacting spins to address individual
spins in an ensemble in the presence of environmental noise, remains an outstanding challenge.

\begin{figure}[t!]
\begin{centering}
\vspace{0cm}
 \hspace{-0.3cm} \includegraphics[width=0.9\columnwidth]{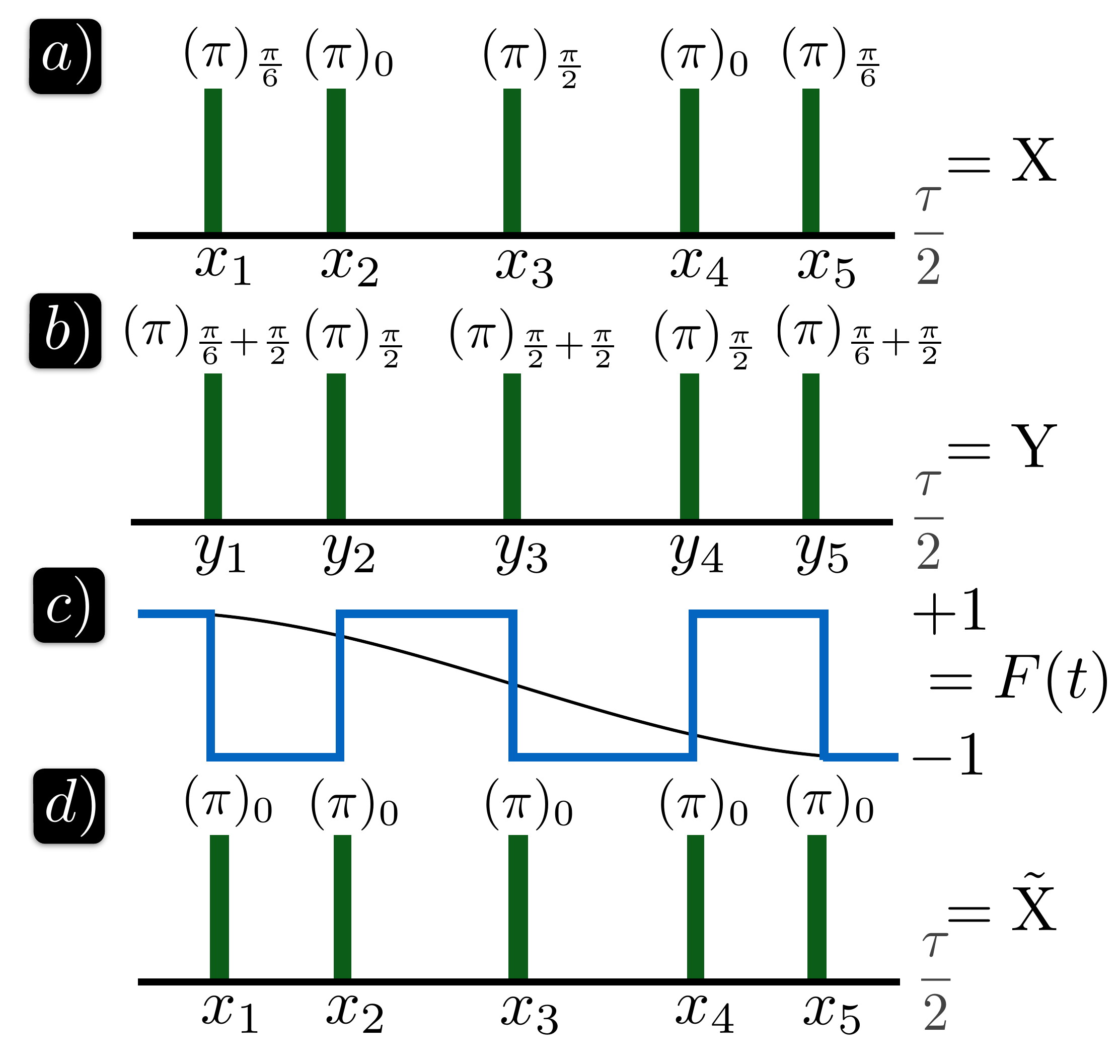}
\par\end{centering}

\caption{(Color online) a) and b) Robust composite pulses with adaptable time variables $x_{i}$
and $y_{i}$  to increase the spectral resolution. Each bar corresponds
to a $\pi$ rotation around the axis defined by $\varphi$ [denoted as $(\pi)_{\varphi}$].
The X, a), and Y, b), composite pulses will be connected
in the robust AXY-n sequence. c) first Fourier component, a half-period,
of the modulation function $F(t)$ for a composite pulse. A whole period $\tau$
corresponds to two successive X and (or) Y composite pulses. d) non-robust
pulse unit $\tilde{\rm X}$ where all $\pi$ rotations are applied in the same~direction}
\label{pulses}
\end{figure}

Here we present a method for the tunable control and selective individual addressing
of nuclear spins via the application of microwave radiation. Our protocol, denoted
adaptive XY-$n$ (AXY-$n$), based on non-equally spaced decoupling pulses \cite{Albrecht2015},
generates highly selective filter functions which ensures a precise discrimination
of the magnetic response of external nuclear spins. We demonstrate the
robustness of our method under realistic conditions including frequency
detunings and amplitude fluctuations of the microwave driving. In this manner, our protocol represents a realistic procedure to achieve high-fidelity quantum gates between the quantum sensor and a spin cluster for quantum computing or quantum simulation purposes. At the same time, this proposal constitutes a highly selective detection method of different nuclear spins in the environment  for structure determination with potential applications in biology as well as solid state physics.

\section{Basic Model}
In the presence of a strong magnetic field $B_{z}$, aligned with
the NV axis $\hat{z}$, the Hamiltonian of the coupled NV center-nuclei system reads
($\hbar=1$)
\begin{eqnarray}
    H = DS_{z}^{2}-\gamma_{e}B_{z}S_{z}-\sum_{j}\gamma_{j}B_{z}I_{j}^{z} +S_{z}\sum_{j}\vec{A}_{j}\cdot\vec{I}_j + H_{\rm c}\label{model}.
\end{eqnarray}
Here $\gamma_{e,j}$ is the electronic, nuclear, gyromagnetic ratio and $H_c$ describes the external microwave control
fields. The interaction between NV center and nuclei is mediated by the
hyperfine vector $\vec{A}_{j}=\frac{\mu_{0}\gamma_{e}\gamma_{j}}{4\pi|\vec{r}_{j}|^{3}}[\hat{z} - 3\frac{(\hat{z}\cdot\vec{r}_{j})\vec{r}_{j}}{|\vec{r}_{j}|^{2}}]$, $\vec{r}_j$ being the vector connecting the NV center and each nucleus in the
environment. Due to the large zero field splitting $D=(2\pi)2.87$ GHz, we have eliminated non-secular components in Eq.~(\ref{model}). For the sake of clarity we neglect dipolar interactions
between nuclei in our analytical work, but retain these interactions
in our numerics. In the rotating frame of  $DS_{z}^{2} - \gamma_{e}B_{z}S_{z}$ and the control field $H_c$ we obtain
\begin{equation}
    H=-\sum_{j}\omega_{j}\ \hat{\omega}_{j}\cdot \vec{I}_{j}+\frac{m_{s}}{2}F(t)\ \sigma_{z}\sum_{j}\vec{A}_{j}\cdot\vec{I}_{j}.\label{casi}
\end{equation}
Here,  $\hat{\omega}_{j}$ is the unit vector  of $\vec{\omega}_j=\gamma_{j} B_{z}\hat{z}-\frac{m_{s}}{2}\vec{A}_j$,
 $\omega_j = |\vec{\omega}_j|$ is the effective Larmor frequency,  and $\sigma_{z} =
|m_{s}\rangle\langle m_{s}|-|0\rangle\langle0|$ the
Pauli operator in the manifold of NV electron spin states $m_{s}=\pm1$ and $m_{s}=0$. For example, when the $m_s=1$  electron spin state is chosen, a qubit between the $m_s = 1$ and $m_s =0$ energy eigenstates is available  while the $m_s=-1$ does not participate in the dynamics governed by Eq.~(\ref{model}). 

\begin{figure}[t!]
\begin{centering}
    \vspace{0cm}
    \hspace{-0.3cm} \includegraphics[width=0.9\columnwidth]{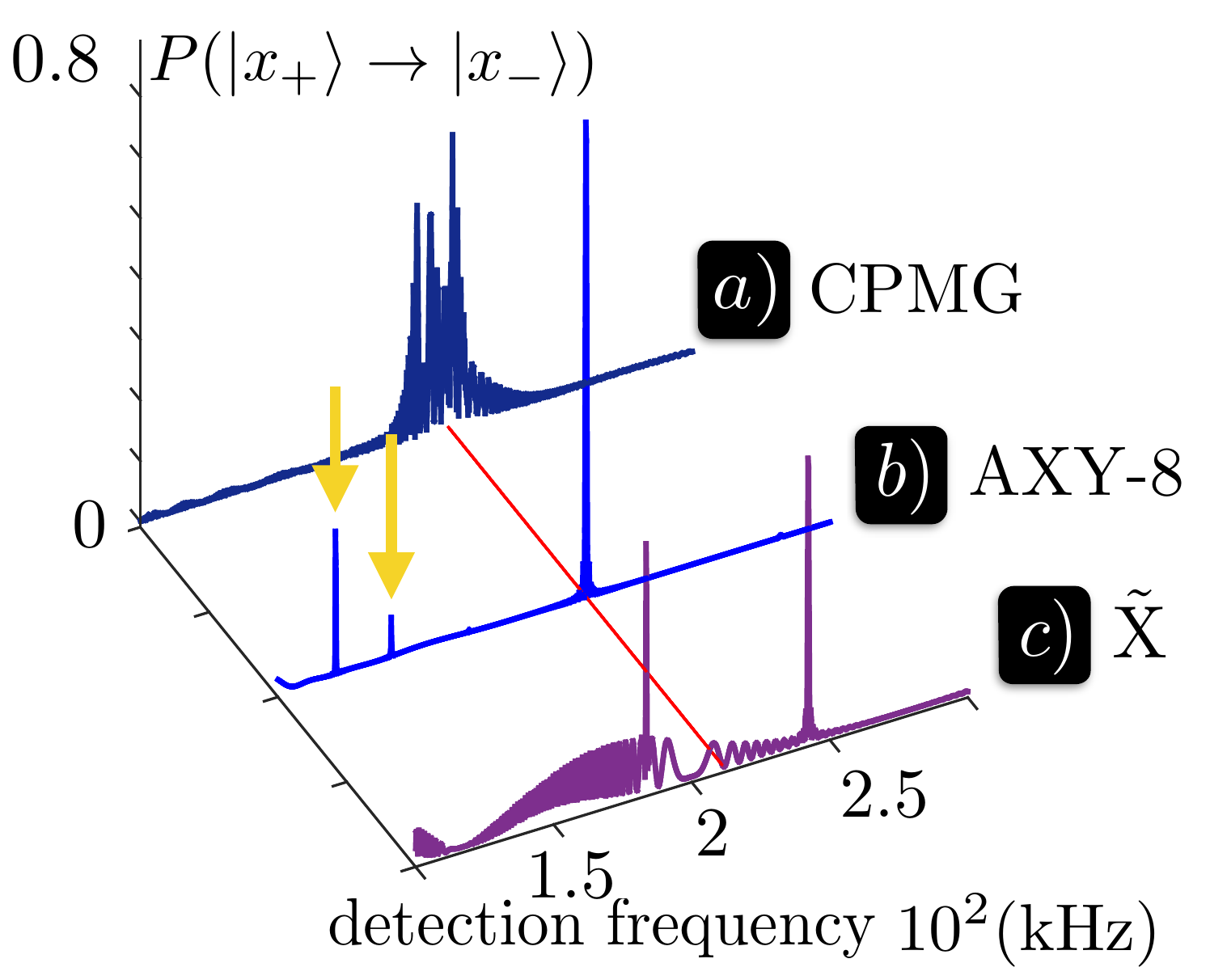}
    \par\end{centering}
    \caption{(Color online) Sensor response (transition probability of an NV center coupled
    to a single nuclear spin) for different pulse sequences. The total evolution time is around
    $1.4$ ms, $B_z=200$ G, and microwave $\pi$-pulses with a duration of $12.5$ ns are applied.
    A detuning of $\Delta=1{\rm MHz}$ and a microwave amplitude mismatch of $5\%$
    have been included. a) response for a CPMG sequence (600 pulses). b) uses
    the AXY-$8$ sequence, while in c)  each X, or Y  is replaced  by the ${\rm \tilde{X}}$
    composite pulse. The patterns in b) and c) are the result of the application of
     $3040$ decoupling pulses with an interpulse spacing such that $f_{1_{DD}} = \frac{1}{10}
    \frac{4}{\pi}$ and $f_{2}=f_{3}=f_{4}=0$. In b) we denote, with arrows, the location of spurious
    resonances, see the main text. The red line crossing the figure denotes the location of
    the resonance frequency.}
\label{patterns}
\end{figure}
The modulation function $F(t)$ takes the values $+1$ ($-1$) depending on whether an even (odd)
number of $\pi$-pulses (instantaneous in our analytical model) have
been applied, see Fig.~\ref{pulses}. The appropriate tuning of $F(t)$ becomes crucial
for accurate spin detection and control. We assume periodic pulse sequences, $F(t)=F(t+\tau)$
so that $F(t)=\sum_{k=0}^{\infty}f_{k}
\cos(k\omega_{DD}t)$, where $\omega_{DD}=2\pi/\tau$ and the Fourier coefficients are
$f_{k}=\frac{2}{\tau}\int_{0}^{\tau}F(t)\cos(k\omega_{DD}t)dt.$
The number of pulses and the spacing between them control $F(t)$ and therefore
the values of the $f_{k}$. For example, CPMG sequences with pulse interval
$\tau/2$ have only two pulses in a DD period $\tau$, resulting in Fourier
coefficients $f_{0}=0$ and $f_{k}=4(k\pi)^{-1} \sin{(\frac{k\pi}{2})}$ for $k>0$.

As the interaction between nuclei and sensor is governed by the second term
of Eq.~(\ref{casi}), it is convenient to move to a rotating frame w.r.t. $-\sum_{j}\omega_{j}\ \hat{\omega}_{j}\cdot \vec{I}_{j}$.
We get
\begin{equation}
H_{{\rm int}} = m_{s}F(t)\frac{\sigma_{z}}{2}\sum_{j}\vec{I}_{j}\cdot\vec{A}_{j}(t),
\end{equation}
where $\vec{A}_{j}(t)= \vec{a}_j\cos(\omega_j t)  +\vec{A}_{j}\times\hat{\omega}_{j}\sin(\omega_j t)+\vec{A}_{j}\cdot\hat{\omega}_{j} \ \hat{\omega}_{j}$
with $\vec{a}_{j}\equiv\vec{A}_{j} -\vec{A}_{j}\cdot\hat{\omega}_{j} \ \hat{\omega}_{j}$.
Now we neglect the last two terms in  $\vec{A}_{j}(t)$ because $F(t)$ is symmetric 
and oscillates with zero mean. Additionally we eliminate the counter
rotating terms of  $H_{{\rm int}}$ finding 
\begin{equation}
H_{{\rm int}} = m_{s}\frac{\sigma_{z}}{8}\sum_{k}f_{k}\sum_{j}\vec{I}_{j}\cdot\vec{a}_{j}[e^{i(\omega_{j}-k\omega_{DD})t}+{\rm c.c.}].
\end{equation}
In the following we identify the conditions that  allow  to tune the
control field in order to address selectively the $n$-th spin. When the
$k_{DD}$-th harmonic matches the Larmor frequency, $k_{DD}\omega_{DD}=\omega_{n}$, we have
\begin{eqnarray}
    H_{{\rm int}} & = & \frac{m_{s}\sigma_{z}}{4}f_{k_{DD}}\vec{I}_{n}\cdot\vec{a}_{n}\nonumber \\
    &  & +\frac{m_{s}\sigma_{z}}{8}\sum_{j\neq n}f_{k_{DD}}\vec{I}_{j}\cdot\vec{a}_{j}\left[e^{i(\omega_{j}-\omega_{n})t}+{\rm c.c.}\right]\label{big}\\
    &  & +\frac{m_{s}\sigma_{z}}{8}\sum_{k\neq k_{DD}}\sum_{j}f_{k}\vec{I}_{j}\cdot\vec{a}_{j}\left[e^{i(\omega_{j}-k\omega_{n}/k_{DD})t}+{\rm c.c.}\right].\nonumber
\end{eqnarray}
For a sufficiently strong magnetic field  $\omega_{j}\sim\gamma_{j}B_{z}\gg|\vec{A}_{j}|$
the last line in Eq.~(\ref{big}) can be ignored if
\begin{equation}
    |\gamma_{j}B_{z}|\gg k_{DD}|\vec{a}_{j}|.\label{eq:largeBz}
\end{equation}
Similarly, the second line in Eq.~(\ref{big}) is eliminated if
\begin{equation}
    |\omega_{j}-\omega_{n}|\gg \big| f_{k_{DD}} \big| |\vec{a}_{j}|. \label{eq:largeAParall}
\end{equation}
For  sequences such as CPMG, XY-type, the Knill DD~\cite{Ryan10,Souza11}, and their combinations~\cite{Souza12,Farfurnik15} where decoupling pulses are equally spaced, the coefficient
$f_{k_{DD}} = 4 \sin{(\frac{k_{DD}\pi}{2})}/(k_{DD}\pi)$ is not tunable and the strong interactions
${\vec a}_j$ limit the addressability because of Eq.~(\ref{eq:largeAParall}). Additionally,
Eq.~(\ref{eq:largeBz}) limits the largest possible $k_{DD}$.

Now we will show how  our approach  can achieve arbitrary tuning of the value
of the coefficient $f_{k_{DD}}$ to obtain the effective Hamiltonian
\begin{equation}
H_{\text{int}}=\frac{m_{s}}{4}f_{k_{DD}}\sigma_{z}\vec{I}_{n}\cdot\vec{a}_{n},
\end{equation}
and therefore individual control of nuclear spins without the necessity of using high harmonics,
i.e. large values of $k_{DD}$.
\begin{figure}[t!]
\begin{centering}
\vspace{0cm}
 \hspace{-1.0cm} \includegraphics[width=1.05\columnwidth]{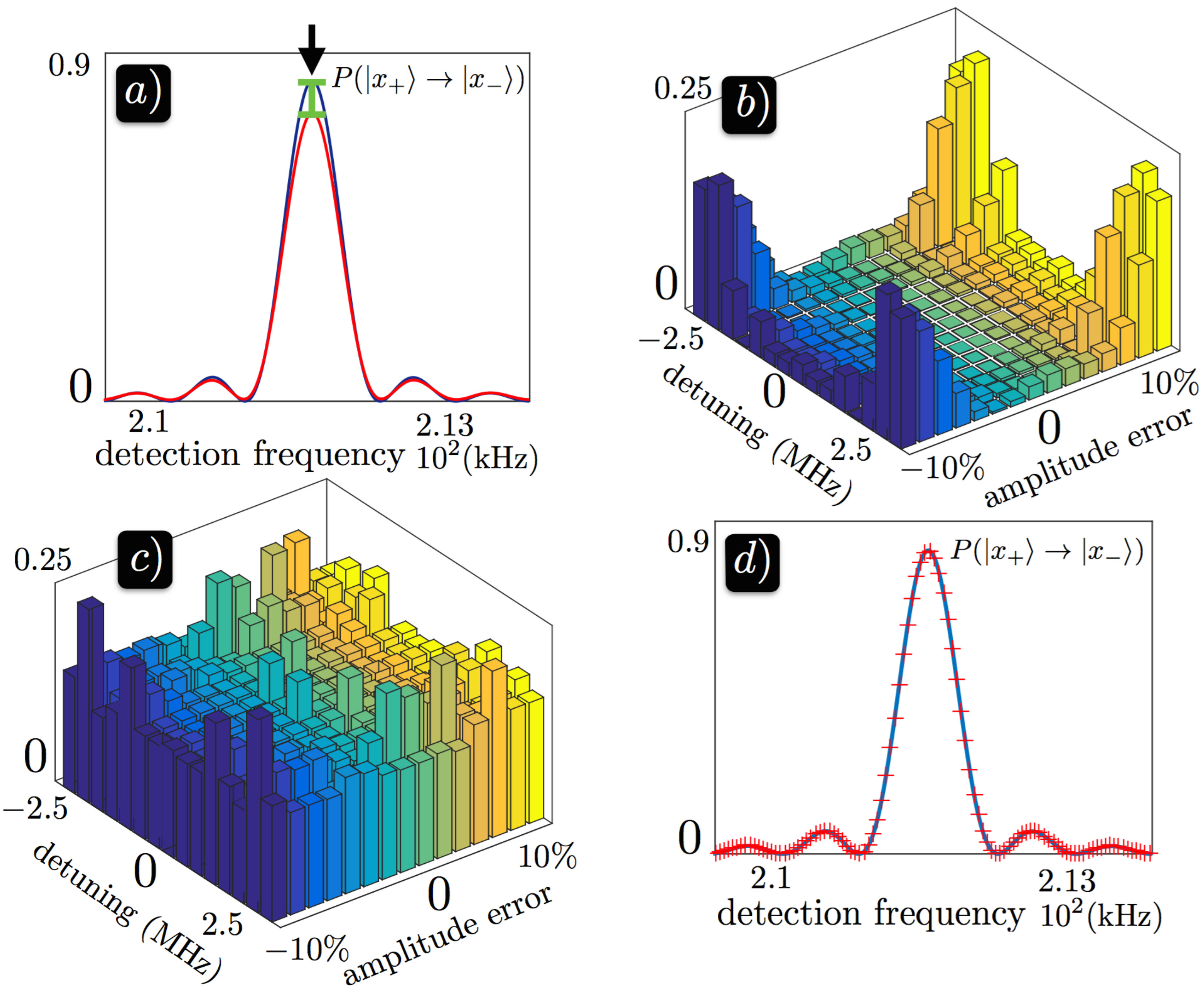}
\par\end{centering}
\caption{(Color online) Difference between ideal and error-included patterns. a) quantity used for comparing the 
behavior of distinct sequences in the presence of errors -
the difference between transition probabilities, see arrow, averaged over the frequency interval
of the figure. In b), c), the difference for the sequences used in Figs.~\ref{patterns} b), c),
is obtained for different values of the detuning and amplitude error. d) shows the patterns without
(solid line) and with (red crosses) time-dependent microwave fluctuation in the control field.}
\label{comparison}
\end{figure}

\section{Shaping the filter functions} 

Since $f_{k}$ depends on the overlap between $F(t)$ and each harmonic component, see Fig.~\ref{pulses} c), we can select sequences
that tune  $f_{k}$ at will. Figs.~\ref{pulses} a) and b) present an example of our proposed
pulse scheme where each X or Y composite pulse is  made of 5 $\pi$-rotations applied along different
directions for enhanced robustness. When equally spaced,  X and Y  converge to
the Knill composite pulse used in the literature~\cite{Ryan10}. Every period $\tau$ includes two
composite pulses X and Y that are applied time symmetric around the midpoint of the period but are
otherwise arbitrary. This guarantees the even character of the Fourier expansion of $F(t)$. Accordingly,
we have $5$ controllable relative time variables $x_{j}$, or $y_{j}$,  each of them measuring the relative positions of the first
5 pulses (applied at $0<x_{j}\tau<\frac{\tau}{2}$). We can use these 5 variables to achieve $f_0=0$,
and $f_{k_{j}^{\prime}}=0$ with $j=1,2,3$ and a tunable coupling constant for the $k$th harmonic
$f_{k}=f_{k_{DD}}$. For example, one can adjust the interpulse spacing in such a way that $f_0=f_2=f_3=f_4 = 0$ and $f_1\neq0$, meaning $k_{1}^{\prime} = 2$, $k_{1}^{\prime} = 3$, $k_{1}^{\prime} = 4$ and $k_{DD} = 1$.  

We note that the symmetric construction $x_{3}=\frac{1}{4}$,
$x_{4}=\frac{1}{2}-x_{2}$, and $x_{5}=\frac{1}{2}-x_{1}$ [see Figs.~\ref{pulses} a) and ~\ref{pulses} b)]
enhances robustness because it is able to cancel first and second order error contributions. This allows to extend our protocol to several thousands of imperfect decoupling pulses, see Appendix.  The symmetric construction yields 
\begin{equation}
f_{k} = \frac{4}{\pi k}\left\{ \sum_{j=1,2}(-1)^{j}\left[(-1)^{k}-1\right]\sin(2\pi kx_{j}) + \sin\left(k\frac{\pi}{2}\right)\right\},
\end{equation}
which vanishes for even $k$. As an example we choose $f_{3}=0$ and $f_{1}=f_{k_{DD}}\equiv f_{1_{DD}}$,
and a solution is 
\begin{equation}
x_{1,2}=\frac{1}{2\pi}\arctan\frac{\pm(3f_{1_{DD}}\pi - 12)w_{1}+\sqrt{3w_{2}}}{\sqrt{6}\sqrt{w_{2}-96f_{1_{DD}}w_{1}\pi\pm w_{1}^{2}\sqrt{3w_{2}}}}
\end{equation}
with the functions $w_{1}=4-f_{1_{DD}}\pi$ and $w_{2}=w_{1}[960-144f_{1_{DD}}\pi-12(f_{1_{DD}}\pi)^{2} +
(f_{1_{DD}}\pi)^{3}]$. Here any $f_{1_{DD}}\pi\in(-8\cos\frac{\pi}{9}+4, 8\cos\frac{\pi}{9}-4)$ is possible. 
Note that for $f_{1_{DD}}\approx0.16$, the solution $x_{1,2}$ reproduces the 5-pulse repetition 
unit in~\cite{Zhao14} (where all $\pi$ rotations are applied about the same axis) which is designed for identifying 
coupled nuclei in a weak field regime {\em and} a semiclassical picture which can 
deviate from a full quantum description~\cite{Zhao11, Huang11}. For equally spaced pulses 
AXY-$n$ converges to the KDD$_{\phi}$ - KDD$_{\phi+\frac{\pi}{2}}$
sequences which are robust against imperfect pulses~\cite{Souza11,Souza12,Farfurnik15}. Additionally, we
can use higher harmonics, $k_{DD}>1$, to reduce the pulse number or to increase the total evolution time at a given magnetic field.
For $f_{1}=0$ and $f_{3}=f_{3_{DD}}$, we obtain 
\begin{equation}
x_{j} = \frac{1}{4} -\frac{1}{2\pi}\arctan\sqrt{q_{j}^{2}-1}, 
\end{equation}
with $q_{j}=4/[\sqrt{5+\pi f_{3_{DD}}}+(-1)^{j}]$ and $j=1,2$, and $f_{3_{DD}}\in(-\frac{4}{\pi},\frac{4}{\pi})$.
\section{Robustness against  errors} 

In Fig.~\ref{patterns} we compare the transition probability
between the $|x_\pm\rangle = \frac{1}{\sqrt{2}} (|0\rangle \pm |1\rangle)$ NV states for different decoupling sequences.
Possible error sources are included in the control Hamiltonian  
\begin{equation}
H_c  = \Omega(t)\cos[ (\omega_{NV} + \Delta) t][S_{x}\cos\varphi(t)+S_{y}\sin\varphi(t)],
\end{equation}
where $\omega_{NV}$ is the resonant frequency of the NV transition, and the microwave pulses are imperfect, with a detuning error $\Delta=1$MHz and a $5\%$ of error in the Rabi frequency $\Omega(t)$. 
The detuning error $\Delta$ takes into account the energy of the nitrogen spin intrinsic to the NV center. Because both the magnetic field and the hyperfine field $\vec{A}_j$ for the nitrogen spin are oriented along the NV symmetry axis, flipping of the nitrogen spin can be neglected by secular approximation. When the intrinsic nitrogen spin is polarized, the detuning $\Delta$ can be made small by calibration. If the  intrinsic nitrogen spin is not polarized, uncertainty of the nitrogen spin state introduce large energy shifts on the splitting between the NV states through the hyperfine interaction, and the unknown energy shifts cause the detuning error within the range of about $1$ MHz for naturally occurring $^{14}$N spin~\cite{Doherty13,Loretz15}.

In Fig.~\ref{patterns} a)  the resonance pattern for a CPMG sequence with $f_{1_{DD}} = 4/\pi$ is plotted. Figs.~\ref{patterns} b) and c)
show the result of a pulse sequence with $f_{1}=f_{1_{DD}} = \frac{1}{10}\frac{4}{\pi}$
and $f_{2}=f_{3}=f_{4}=0$. 
While the spectrum in Fig.~\ref{patterns} a) is distorted because of accumulated errors,  Fig.~\ref{patterns} b) shows a clear pattern obtained by the AXY$-8$  sequence (AXY-$n$ for $n=8$, i.e. XYXYYXYX), where the pulses are repeatedly applied according to the pattern of ${\rm  XY-}8$. This is a sequence containing $8$ composite pulses,  X or Y, where X and Y are designed according to the scheme in  Figs.~\ref{pulses} a) and b) respectively. Our simulations extend to $3040$ pulses corresponding to a final evolution time $\approx 1.4$ ms. We observe that the resonant peak is very sharp which is of considerable benefit for the filtering of noisy
signals. Note that common to all pulsed schemes are spurious resonances due to the finite width
of the microwave $\pi$-pulses~\cite{Loretz15}. We would like to stress however that the composite
pulse structure of the AXY-8 sequence produces very narrow spurious peaks which, additionally, are far away from the principal resonance even in the presence of noise and imperfections
(see Fig.~\ref{patterns} b)).
Fig.~\ref{patterns} c) has been computed  repeating the $\tilde{X}$
sequence, see Fig.~\ref{pulses} d), where all  $\pi$ rotations are applied along the
same axis. We can observe how the accumulated errors distort the pattern destroying any
readable~signal. 

To investigate the behavior of each multipulse sequence in  presence of strong error
sources we compare the different patterns that appear when changing the conditions in $H_c$.
In Fig.~\ref{comparison} a) the cases for $\Delta=0$ and perfect microwave pulses (blue line), and
$\Delta=1$ MHz and a $10\%$ of error in the rotating Rabi frequency (red line) for the  the AXY$-8$
sequence are plotted. In Figs.~\ref{comparison} b) and c) we show the average difference
between the transition probabilities in the ideal case and in the presence of error sources in a region
close to the resonance position. The difference between patterns is small for AXY-$8$ in a
detuning range of $\pm1.5$ MHz and $\pm10\%$ in the microwave driving error, see
Fig.~\ref{comparison} b). In the case of the combined sequence $\tilde{X}$, Fig.~\ref{comparison} c),
the pulse errors  rapidly accumulates. In this manner we confirm that the AXY$-8$  sequence is robust against realistic error sources allowing, at the same time, to accurately filter specific
frequency-components.

\begin{figure}[t!]
\begin{centering}
\vspace{0cm}
 \hspace{-0.95cm} \includegraphics[width=1.0\columnwidth]{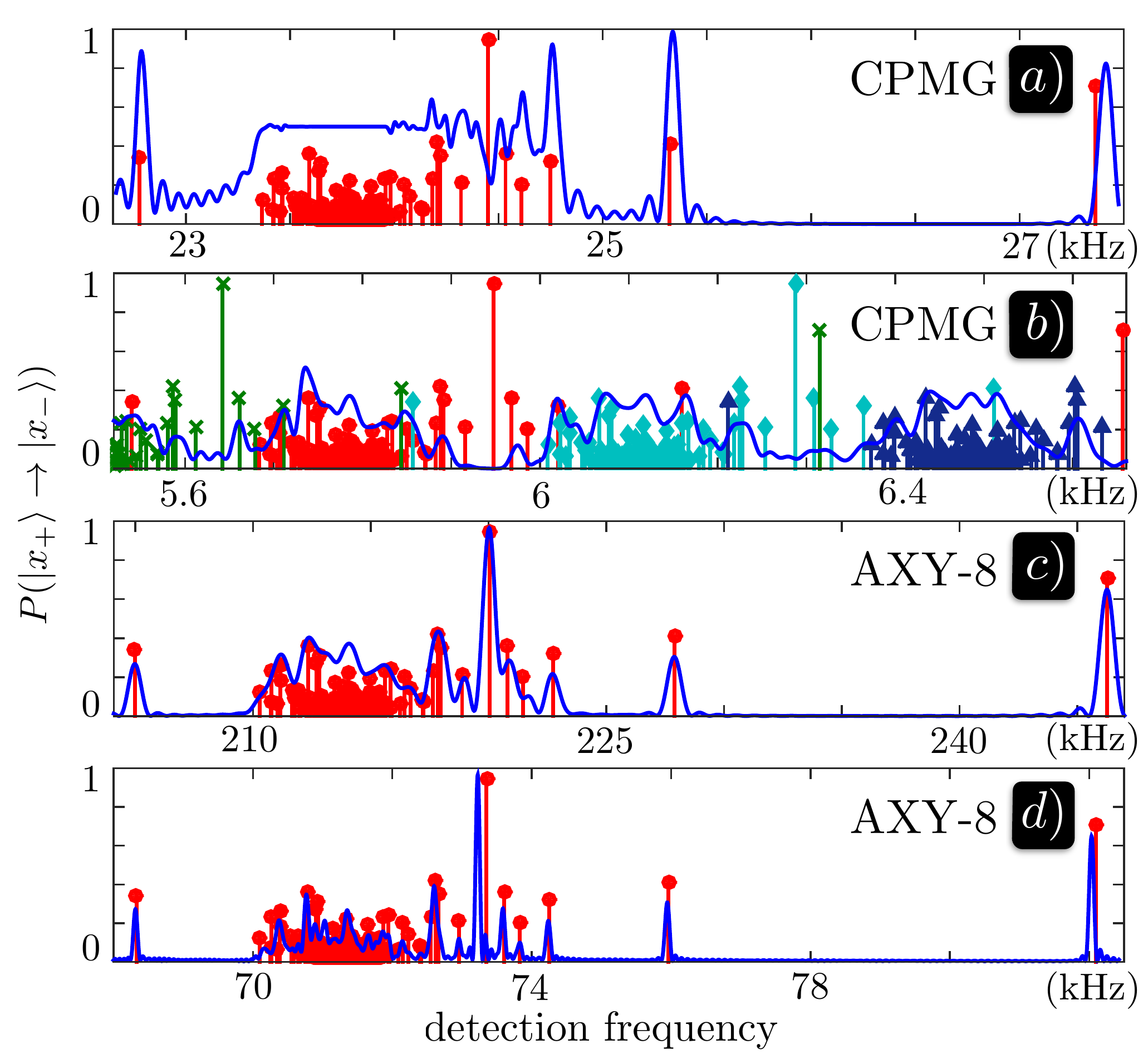}
\par\end{centering}
\caption{a) (Color online) Resonance pattern for a CPMG sequence, $9^{th}$ harmonic
used. The position of the vertical lines denote  $\omega_j/k$ and their height
denotes the strength of interaction between  the sensor and the nucleus. In b) we can observe the effect
of the overlap caused by $k=35$ (diamonds), $33$ (triangles), and $39$ (crosses) harmonics
on the spectrum of the $k=37$ (circles) harmonic. c) and d) show resonance pattern
in the AXY-$8$  sequence for the $1^{st}$ and $3^{rd}$ harmonics respectively. In each
plot, the horizontal axis corresponds to the frequency range of the modulation function
one should explore in order to obtain the nuclear response in the appropriate harmonic.}
\label{final}
\end{figure}

Our AXY-$8$ is also robust against amplitude fluctuations in the driving field. In Fig.~\ref{comparison} d) we compare the result with (red crosses) and  without (solid line) fluctuations in the control field. Fluctuations are
modeled by an Ornstein-Uhlenbeck processes
where the amplitude fluctuation $\Delta\Omega(t)$ evolves according to $\Delta\Omega(t+\Delta t)
= \Delta\Omega(t)e^{-\Delta t/\tau_{{\rm mw}}} + n_{{\rm G}}\sqrt{\frac{c_{{\rm mw}}\tau_{{\rm mw}}}{2}
(1-e^{-2\Delta t/\tau_{{\rm mw}}})}$~\cite{Gillespie96},  $\tau_{{\rm mw}}\approx 1$ ms being the correlation time of the
microwave noise and $c_{{\rm mw}}\approx 2\delta^2_\Omega \Omega^2 \tau_{{\rm mw}}^{-1}$
with the relative amplitude fluctuation $\delta_\Omega \approx 7 \times10^{-3}$,
see Ref~\cite{Cai12}. Fig.~\ref{comparison} d)  demonstrates that the effect of  microwave
fluctuations is negligible in the resonance pattern which confirms the robustness of our protocol.

\section{Single spin addressing}
Finally, we have performed simulations in a diamond containing $736$ $^{13}$C spins
which are generated on random lattice sites of the diamond structure according to the
natural abundance of $1.1\%$ and  $B_z=200$G.  In the simulation we use instantaneous pulses and a disjoint cluster expansion  including dipolar interactions up to a group size of six $^{13}$C~\cite{Maze08bis}. In Figs.~\ref{final} a) and b)
the patterns were obtained by CPMG sequences with a total evolution time $\sim1.4$ ms.
Fig.~\ref{final} a) shows the result for the $9^{th}$ harmonic of the $66$-pulse CPMG sequence.
In this sample, working at higher harmonics causes spectral overlap and difficulty in
distinguishing signals. This can be observed in Fig.~\ref{final} b) where the $37^{th}$
harmonic (hence $16$ pulses) has been used. Using the AXY$-8$  sequence, for the first
[Fig.~\ref{final} c)] and third [Fig.~\ref{final} d)] harmonic, more spins can be resolved
by individual resonant peaks. The sequence in Fig.~\ref{final} c) has been tuned to
$f_{1_{DD}} = \frac{4}{ 37 \pi}$ and $f_{k} = 0$, $k$ being $3$ or any even number,
and it uses $3000$ decoupling pulses to obtain a total time duration equaling that
of Figs.~\ref{final} a) and b). In Fig.~\ref{final} d), the third harmonic version with $f_{3_{DD}}
= \frac{4}{ 111 \pi}$ and $f_k = 0$, $k$ being $1$ or any even number, can provide $3$-fold
evolution time (i.e., $4.2$ ms) compared to the first harmonic version and hence narrower
signal peaks. In both cases the resolution is increased in a way that
permits individual addressing of a considerable number of spins without the problem of spectral overlaps.

\section{Conclusions} We have presented a protocol that allows for individual addressing
of nuclear spins even in large assemblies and when weakly coupled to a sensor. Based on
the application of non-uniformly spaced decoupling pulses our method obviates the need for
high harmonics which reduces significantly peak pollution and
consequently enhances both resolution and robustness. The method is generic as it permits
the high-resolution detection of arbitrary magnetic moments by means of a quantum sensor.
This scheme promises applications in the control of mesoscopic nuclear spin based quantum
registers in diamond as well as the precise  nuclear positioning
in external probes including biomolecules.

\section{Acknowledgements} This work was supported by the Alexander von Humboldt
Foundation, the ERC Synergy grant BioQ, the EU projects DIADEMS, SIQS and
EQUAM as well as the DFG SFB TRR/21. 

J. C. and Z.-Y. W. contributed equally to this work. 

\section{Appendix: Robustness of symmetric AXY-${\bf 8}$}
\subsection{Control errors}
To show the robustness of our protocol for leading and second order error contributions  on the electron spin rotations  it is sufficient to consider the  pulse Hamiltonian 

\[
H=\Delta\frac{\sigma_{z}}{2}+\Omega(t)\frac{\sigma_{\phi(t)}}{2},
\]
where $\Delta$ is a detuning, $\Omega(t)=\Omega$ during the pulse
control and $\Omega(t)=0$ between pulses. The phase $\phi(t)$ of
the Pauli operator

\begin{equation}
\sigma_{\phi}\equiv\cos\phi \ \sigma_{x}+\sin\phi \ \sigma_{y},
\end{equation}
switches for different pulses. One can demonstrate that, up to second order on small parameters $\epsilon = \frac{\Delta}{\Omega}$ and $\delta$, the latter being the microwave amplitude mismatch, the propagator associated to the imperfect rotation is 
\begin{equation}\label{simple}
R_{\phi} = e^{-i\frac{(\pi - 2 \delta)}{\Omega} H} \approx-i \bigg(1 -\frac{\delta^2 + \epsilon^2}{2} \bigg)\ \sigma_{\phi} - i \epsilon \sigma_z + \bigg(\delta -\frac{\epsilon^2\pi}{4}\bigg) \ \mathbb{I}.
\end{equation}
Note that the small quantities $\delta$ and $\epsilon$ can be always written as $\delta = \eta \tilde{\delta}$ and $\epsilon = \eta \tilde{\epsilon}$ with $\eta < 1$, therefore we can make the expansion in terms of the $\eta$ parameter. 
\subsection{Robust composite pulses}
\subsubsection{No pulse delay}
The simplest case is that the pulses are applied subsequently without
delay. A sequence of composite pulses for the X pulse has the propagator
(without pulse delays) X$=R_{\frac{\pi}{6}}R_{0}R_{\frac{\pi}{2}}R_{0}R_{\frac{\pi}{6}}$.
Using Eq. (\ref{simple})  and keeping up to the leading order (first order)
of errors we find 
${\rm X} = O_0+O(\eta^{2})$,
where
\begin{equation}
O_0 =  i\sigma_x e^{-i\frac{\sigma_z}{2}\frac{\pi}{3}},
\end{equation}
is the perfect rotation for the X composite pulse. Therefore the  X composite pulse  cancels the leading order error.

\subsubsection{Including pulse delay}
We include pulse delays $t_i$ between the $(i-1)$-th and  $i$-th pulses.
In the interpulse spacing, the presence of  the detuning $\Delta$  cannot be neglected because the accumulated effect $\Delta t_i$ is not small for long pulse delays.
With the pulse delays,  the propagator of the imperfect composite X pulse is 
\begin{equation}\label{big1}
{\rm X} = U_{\frac{\pi}{6} \Delta_5} U_{0 \Delta_4} U_{\frac{\pi}{2} \Delta_3} U_{0 \Delta_2} U_{\frac{\pi}{6} \Delta_1}, 
\end{equation}
where 
\begin{eqnarray}\label{condition}
\Delta_1 &\equiv& 2\Delta (t_1),\nonumber\\
\Delta_2 &\equiv& 2\Delta (t_2- t_1),\nonumber\\
\Delta_3 &\equiv& 2\Delta (t_3+ t_1 - t_2),\nonumber\\
\Delta_4 &\equiv& 2\Delta (t_4+ t_2 - t_3 - t_1),\nonumber\\
\Delta_5 &\equiv& 2\Delta (t_5+ t_3 + t_1 - t_4 - t_2 ),
\end{eqnarray}
and 
\begin{equation}\label{propagator}
U_{\phi,\Delta_i} = i(\eta^2 \tilde{\alpha}^2 - 1)\sigma_{\phi} - i \eta\tilde{\epsilon} \sigma_z U_{\Delta_i} + \eta\tilde{\delta} U_{\Delta_i} -\eta^2\tilde{\beta}^2 U_{\Delta_i},
\end{equation}
with $\tilde{\alpha}^2 = \frac{\tilde{\delta}^2 + \tilde{\epsilon}^2}{2}$, $\tilde{\beta}^2 = \frac{\tilde{\epsilon}^2\pi}{4}$, and $U_{\Delta_i} = e^{-i\Delta_i\sigma_z}$. The imperfect composite Y pulse is related with X pulse by
\begin{equation}\label{big2}
{\rm Y} =  U_{z}(\pi/2) \ {\rm X} \ U_{z}^\dag(\pi/2),
\end{equation}
with $U_{z}(\pi/2) = e^{-i \frac{\sigma_z}{2} \frac{\pi}{2}}$.
Now Eq.~(\ref{big1}) can be expanded in different orders in $\eta$.  For symmetric sequences we have 
\begin{eqnarray}\label{symmetry}
&\Delta_{4} + \Delta_{5} = \Delta_{2} +\Delta_{1},\nonumber\\
&\Delta_{4} = \Delta_{2},\nonumber\\
&\Delta_{5} = \Delta_{1}.
\end{eqnarray}
One can obtain the first order correction
\begin{equation}\label{first}
O_1= \eta (i \tilde{\epsilon} \sigma_z - \tilde{\delta}) (U_{\Delta_1} - U_{\Delta_3}).
\end{equation}
When the pulses are equally spaced we have $t_3 = t_2$ which according to Eqs.~(\ref{condition}) implies  $\Delta_1 =\Delta_3$ and therefore $U_{\Delta_1} - U_{\Delta_3}= 0$  which leads to $O_1= 0$. To suppress higher order errors we will combine the composite X and Y pulses in XY sequences. Additionally we will demonstrate that even when the pulses on the inner level of the X or Y sequences are not equidistant but distributed according to Eq.~(\ref{symmetry})  the errors up to second order in $\eta$ can be cancelled.

\subsubsection{{\rm AXY}-$4$ sequence}
When  $\Delta_1 \neq \Delta_3$, that is the pulses are not equally spaced,  $O_1\neq 0$. Now we show how leading order error cancellation  appear because  the combination of the composite X and Y pulses, Eqs.~(\ref{big1}) and (\ref{big2}), in the  XY-$4$ configuration (this scheme constitutes the AXY-$4$ sequence).  
Up to the first order in $\eta$  we have
\begin{equation}\label{xpulse}
{\rm X} = O_0  +   O_1,  
\end{equation}
\begin{equation}\label{ypulse}
{\rm Y}=   U_{z}(\pi/2)(O_0  +  O_1 ) U^\dag_{z}(\pi/2),
\end{equation}
Now, by simply calculating the first order contribution in $\eta$ to AXY-$4$ sequences (here XYXY, or YXYX)  we have that for the XYXY sequence that term is
\begin{equation}
O_0 O^z_0  (O_0 O_1^z + O_1 O_0^z) + (O_0 O_1^z + O_1 O_0^z) O_0 O^z_0  = 0,
\end{equation}
while for the YXYX reads
\begin{equation}
O^z_0 O_0  (O^z_0 O_1 + O^z_1 O_0) + (O^z_0 O_1 + O^z_1 O_0) O^z_0 O_0   = 0,
\end{equation}
where  $O^z_j = U_{z}(\pi/2)  O_j U_{z}(\pi/2)^\dag$. This means that terms going with $\eta$ are cancelled because of the XYXY, or YXYX, construction. However, one can demonstrate that  this cancellation does not require  the application of the conditions in Eq.~(\ref{symmetry}).

\subsubsection{{\rm AXY}-${8}$ sequence}
The symmetry imposed by our protocol, i.e. the conditions in Eq.~(\ref{symmetry}), allows us to eliminate the second-order error contribution when  we combine our composite pulse with the XY-8 sequence, namely, the AXY-$8$ protocol is applied. Note that  to cancel the second-order error symmetric composite pulses are required. Non-symmetric composite pulse only eliminate the the leading order error because of the reasons explained in the previous subsection. To make the argument complete let us additionally introduce second order  error contributions in the discusion. This reads
\begin{eqnarray}
O_2 &=& -i \eta^2(\tilde{\epsilon}^2 + \tilde{\delta}^2) (\frac{3}{2} \sigma_{\frac{\pi}{6}} \sigma_y   \sigma_{\frac{\pi}{6}} + \sigma_y) +\eta^2 \tilde{\beta}^2 (U_{\Delta_{1}} - U_{\Delta_{3}})\nonumber\\ 
&&+i \eta^2 ( \tilde{\delta}^2 - \tilde{\epsilon}^2) (U_{\Delta_{3+2}} \sigma_{\frac{\pi}{6}}\sigma_{x}\sigma_{\frac{\pi}{6}} + \sigma_{\frac{\pi}{6}}\sigma_{x}\sigma_{\frac{\pi}{6}} U_{\Delta_{3+2}}  )\nonumber\\
&&+2 \eta^2(\tilde{\epsilon} \tilde{\delta}) (U_{\Delta_{3+2}} \sigma_{\frac{\pi}{6}}\sigma_{x}\sigma_z\sigma_{\frac{\pi}{6}} + \sigma_{\frac{\pi}{6}}\sigma_z\sigma_{x}\sigma_{\frac{\pi}{6}} U_{\Delta_{3+2}}  )\nonumber\\
&& + i \eta^2 (\tilde{\epsilon}^2 + \tilde{\delta}^2) (U_{\Delta_{1-3}}\sigma_{\frac{\pi}{6}} + \sigma_{\frac{\pi}{6}}U_{\Delta_{1-3}}  ).
\end{eqnarray}
Therefore, up to the second order one can write
\begin{equation}\label{xpulse}
{\rm X} = O_0  +   O_1 +   O_2, 
\end{equation}
\begin{equation}\label{ypulse}
{\rm Y}=   U_{z}(\pi/2)(O_0  +  O_1 +  O_2 ) U^\dag_{z}(\pi/2).
\end{equation}

Now under the symmetry conditions in  Eq.~(\ref{symmetry}) we have up to the second order of errors
${\rm XYXYYXYX} = (\tilde{f}_0 + \tilde{f}_2) (f_0 + f_2)$, where 
\begin{eqnarray}
&\tilde{f}_0 = O_0 O^z_0 O_0 O^z_0,\nonumber\\
&f_0 = O^z_0 O_0 O^z_0 O_0,\nonumber\\
&\tilde{f}_2 = O_0 O^z_0 (O_0 O_2^z +O_1 O_1^z + O_2 O_0^z )\nonumber\\
& + (O_0 O_2^z +O_1 O_1^z + O_2 O_0^z )O_0 O^z_0\nonumber\\
& + (O_0 O_1^z + O_1 O_0^z) (O_0 O_1^z + O_1 O_0^z), \nonumber\\
&f_2 = O^z_0 O_0 (O^z_0 O_2 +O^z_1 O_1 + O^z_2 O_0 )\nonumber\\
& + (O^z_0 O_2 +O^z_1 O_1 + O^z_2 O_0 ) O^z_0 O_0\nonumber\\
& + (O^z_0 O_1 + O^z_1 O_0) (O^z_0 O_1 + O^z_1 O_0). \nonumber\\
\end{eqnarray}
It can be calculated that the errors up to the second order $\tilde{f}_0 f_2 + \tilde{f}_2 f_0 = 0$  is eliminated by the symmetric contruction of the AXY-$8$ sequences. That is, we have the robust sequence 
\begin{equation}
{\rm AXY-}8 = {\rm XYXYYXYX} = \mathbb{I} + O(\eta^3),
\end{equation} 
where we have used that the error-free control $ \tilde{f}_0 f_0 = \mathbb{I}$.

\end{document}